\documentclass[12pt]{article}
\usepackage[latin1]{inputenc}
\usepackage{amsfonts}
\usepackage{amssymb,amsmath}

\def\ep{\varepsilon}

\def\diag{\mathrm{diag}}
\def\p{\partial}

\def\GG{\mathrm{G}}
\def\da{{\dot a}}
\def\db{{\dot b}}
\def\ep{\varepsilon}
\def\mn{{\mu\nu}}

\def\({\left\{}
\def\){\right\}}
\def\[{\left[}
\def\]{\right]}

\def\Pg{{\mathbb{P}_\mathrm{G}}}

\def\Proy{\mathbb{P}}

\newcommand{\cor}[1]{\left\{#1\right\}}
\newcommand{\dir}[1]{\left\{#1\right\}_{\textrm{D}}}
\newcommand{\con}[1]{\left[#1\right]}
\newcommand{\dud}[3]{_{#1\phantom{#2}#3}^{\phantom{#1}#2}}
\newcommand{\udu}[3]{^{#1\phantom{#2}#3}_{\phantom{#1}#2}}
\newcommand{\du}[2]{_{#1}^{\phantom{#1}#2}}
\newcommand{\ud}[2]{^{#1}_{\phantom{#1}#2}}
\newcommand{\mat}[1]{\left(\begin{matrix}#1\end{matrix}\right)}

\title{ Quantum algebra of $N$ superspace }  \vskip 5.5cm
\author{Nicol{\'a}s Hatcher\\ \vspace{2 cm}\\ A. Restuccia, J. Stephany}
\date{}
\begin{document}

\maketitle \thispagestyle{empty} \vspace{-10cm} {\hfill{Preprint
{\bf SB/F/06-337}} \hrule \vspace{5.5cm}
\begin{center}
\textit{ Universidad Sim{\'o}n Bol{\'\i}var, Departamento de Matem{\'a}ticas, \\
Apartado Postal 89000, Caracas 1080-A, Venezuela.}\vspace{2cm}\\
\textit{Universidad Sim{\'o}n Bol{\'\i}var, Departamento de F{\'\i}sica\\
Apartado 89000, Caracas 1080A, Venezuela.}\vspace{1.5cm}\\
\textit{nhatcher@fis.usb.ve, arestu@usb.ve, stephany@usb.ve}
\end{center}
\begin{abstract}

We identify the quantum algebra of position and momentum operators
for a quantum system  bearing an irreducible representation of the
super Poincar{\'e} algebra in the $N>1$ and $D=4$ superspace both in
the case where there are not central charges in the algebra and
when they are present. This algebra is noncommutative for the
position operators. We use the properties of superprojectors
acting on the superfields to construct explicit position and
momentum operators satisfying the algebra. They act on the
projected wave functions associated to the various supermultiplets
with defined superspin present in the representation. We show that
the quantum algebra associated to the massive superparticle
appears in our construction and is described by a supermultiplet
of superspin 0. This result generalizes the construction for
$D=4$, $N=1$ reported recently. For the case $N=2$ with central
charges we present the equivalent results when the central charge
and the mass are different. For the $/kappa$-symmetric case when
these quantities are equal we discuss the reduction to the
physical degrees of freedom of the corresponding superparticle and
the construction of the associated quantum algebra.
\end{abstract}
\vskip 2cm \hrule
\bigskip
\centerline{\bf UNIVERSIDAD SIM{\'O}N BOL{\'I}VAR} \vfill

\section{Introduction}
The  adequate formulation  of the  quantum mechanics for a given
classical system has been the subject of much attention ever since
the method of canonical quantization was proposed in the late
twenties but there are still unanswered questions in relation to
this important problem. In the presence of second class
constraints, neither the Dirac procedure nor any of the modern
approaches, including the Becchi-Rouet-Stora-Tyutin (BRST)
construction, give a general recipe to represent the algebra of
observables. It has been proposed \cite{BatF1983,BizS1996} that
any second class constrained system may be extended by introducing
new variables to a first class constrained one such that under
partial or total gauge fixing reduces to it. But although there
are cases in which this approach provides a solution to the
construction of the observables of the theory, there are others
for which it does not. In particular, for extended systems, it is
not known how to realize this approach locally in a systematic
way. In this letter we discuss this fundamental aspect of the
quantization procedure for superparticles in $D=4$. We show that
properties of the superprojectors which reduce the superfields to
the sectors associated to the irreducible supermultiplets
determine the algebra of the projected position and momentum operators. For the
superspin 0 sector this algebra coincides with the quantum algebra of the massive superparticle. \\

A noncommutative structure of the geometric space  appears in
various physical contexts. The Seiberg-Witten  \cite{SeiW199},
limit of constant antisymmetric $B$ field in string theory leads
to a noncommutativity of the related Super Yang Mills and Super
Born-Infeld fields theories. Besides interacting super D-branes
theories requires noncommutative coordinates expressed as Lie
algebra valued coordinates for consistency. Also the quantization
of the massive spinning  and supersymmetric particles may be
represented in terms of
to a noncommutative superspace. \\

As we review below, for massive superparticles the classical Dirac
algebra of the position operators describe a non commutative
superspace. The Dirac brackets of the $X^\mu$ coordinate variables
are proportional to the internal angular momentum. For the
quantization of this system a set of operators different of the
standard multiplicative position operators which realizes the
noncommutative algebra should be found.  In \cite{ResHS2006} we
found indeed the general solution for the quantum algebra of $N=1$
superparticles in 4 dimensions in terms of the superprojectors  to
the chiral and anti-chiral and tensorial multiplets. In this case
the chiral and anti-chiral $N=1$ multiplets are associated to the
usual $N=1$ superparticle and we found that the $N=1$ tensorial
spin $1/2$ multiplet should be associated to a new superparticle
action \cite{ResHS2006,Hat}.
\\

In this paper we extend our previous work for the $N>1$ and $D=4$
superspace both in the case where there are not central charges in
the algebra and when they are present. After a  discussion of the
classical mechanics of the corresponding massive superparticles
which also serves to identify the general form of the quantum
algebra in each of the  superspaces, we present the explicit
solution for the quantum operators using the superprojectors and
demonstrate that they satisfy the right algebra.\\

We notice that the solution of the quantum algebra is of course
Lorentz covariant, providing a general approach for the covariant
quantization of massive superparticles.

\section{The classical system without central charges}

Let us consider first the massive superparticle
\cite{ResHS2006,Cas2} in $D=4$, $N$, superspace without central
charges (in this section we follow the discussion in
\cite{ResHS2006}). The metric signature is
$\eta_\mn=\diag\{-1,+1,+1,+1\}$ and the superspace coordinates are
$(x^\mu,\theta^{ai},\bar\theta\ud{\da}{i})$, where $a=1,2$ is a
spinor index and $i=1,...,N$ is the number of supersymmetric
charges. Naturally $(\theta^{ai})^*=\bar\theta\ud{\da}{i}$. We
choose Dirac matrices to be off-diagonal and given by,
\begin{gather}
\gamma^\mu=\mat{0 & \sigma^\mu\\ \bar \sigma^\mu & 0}
\end{gather}
with $\sigma\ud{\mu}{a\db}$ the Pauli matrices. The action
principle  for the massive superparticle is given by \cite{Cas2}
\begin{gather}
S=\frac{1}{2}\int
d\tau\left(e^{-1}\omega^\mu\omega^\nu\eta_\mn-em^2\right) \ \ ,
\label{SuperAction}
\end{gather}
where  $\omega^\mu=\dot x^\mu-i\dot\theta^{ai}
\sigma\ud{\mu}{a\db}\bar\theta\ud{\db}{i}+i\theta^{ai}\sigma\ud{\mu}{a\db}
\dot{\bar\theta}\ud{\db}{i}$ is defined for convenience. The
generalized momenta are given by
\begin{gather}
\pi_e=0\\
p_\mu=e^{-1}\omega_\mu\\
\pi_{ai}=-ip_\mu \sigma\ud{\mu}{a\db}\bar\theta\ud{\db}{i}\\
\bar\pi\du{\db}{i}=-ip_\mu\bar\theta^{ai}\sigma\ud{\mu}{a\db}\ \ ,
\end{gather}
and satisfy the canonical Poisson bracket relations \cite{Cas2}
\begin{gather}
\cor{x^\mu,p_\nu}=\delta\ud{\mu}{\nu}\\
\cor{\theta^{ai},\pi_{bj}}=-\delta\ud{a}{b}\delta\ud{i}{j}\ \ .
\end{gather}
The total angular momentum  is given by
\begin{gather}
J_\mn=L_\mn+S_\mn\\
L_\mn=x_\mu p_\nu-x_\mu p_\nu\\
S_\mn=-\frac{1}{4}\left(\theta^{ai}{\sigma_\mn}\du{a}{b}\pi_{bi}+
\bar\pi\du{\da}{i}{\bar\sigma_\mn}\,\ud{\da}{\db}\bar\theta\ud{\db}{i}\right)\
\label{ClassicalInternal} \ .
\end{gather}
For this system, there is  one first class constraint $\pi_e=0$
related to  the  reparametrization invariance of the action which
implies the secondary first class  constraint $p^2+m^2=0$. There
also appear the  constraints
\begin{gather}
\label{eq:d1}
d_{ai}\equiv \pi_{ai}+ip_\mu \sigma\ud{\mu}{a\db}\bar\theta\ud{\db}{i}=0 \\
\bar d\du{\db}{i}\equiv \bar\pi\du{\db}{i}+ip_\mu \theta^{ai}\sigma\ud{\mu}{a\db}=0\label{eq:d2}\\
\cor{d_{ai},\bar d\du{\db}{j}}=-i2\delta\du{i}{j}p_\mu
\sigma\ud{\mu}{a\db}\equiv {C_{ai\db}}^j\ \ ,
\end{gather}
which are second class since the matrix ${C_{ai\db}}^j$ is non
singular if $m\neq 0$.  The Dirac brackets are given by
\begin{gather}
\dir{F,G}=\cor{F,G}-\cor{F,d_{ai}}\hat C\ud{ai\db}{j}\cor{\bar
d\du{\db}{j},G}-\cor{F,\bar d\du{\db}{j}}\hat
C\ud{ai\db}{j}\cor{d_{ai},G}\ \ .
\end{gather}
Here  $\hat C\ud{ai\db}{j}=\dfrac{1}{2ip^2}p_\mu\bar
\sigma^{\mu\,\da b}\delta\ud{i}{j}$ and verifies
\begin{gather}
{C_{ai\db}}^j \hat C\ud{bk\db}{j}=\delta\du{i}{i}\delta\du{a}{b}\
\ .
\end{gather}
Calculating the Dirac brackets for all coordinates and momenta the
result is \cite{Cas2}
\begin{gather}
\dir{\theta^{ai},\theta^{bj}}=\dir{\pi_{ai},\pi_{bj}}=\dir{p_\mu,p_\nu}=0\label{con1}\\
\dir{\theta^{ai},\bar\theta\ud{\da}{j}}=\frac{-1}{2ip^2}p_\mu\bar \sigma^{\mu\,\da a}\delta^i_j\\
\dir{x^\mu,\theta^{ai}}=\frac{1}{2p^2}p_\nu \bar\sigma^{\nu\,\da a}\theta^{bi} \sigma\ud{\mu}{b\da}\\
\dir{x^\mu,\bar\theta\ud{\da}{i}}=\frac{1}{2p^2}p_\nu \bar
\sigma^{\nu\,\da a}
\bar\theta\ud{\db}{i}\sigma\ud{\mu}{a\db}\label{con3}\\
\dir{x^\mu,x^\nu}=\frac{-S^\mn}{p^2}\label{con2}\ \ .
\end{gather}
The nontrivial  Poisson bracket (\ref{con2}) was first discussed
by Pryce \cite{Pry} and in the context of superparticles by
Casalbuoni \cite{Cas2}. As we discuss below it is a most important
ingredient in the formulation of the quantum theory.  Note that
using (\ref{eq:d1}) and (\ref{eq:d2}), the equation for the
internal angular momentum $S^\mn$ (\ref{ClassicalInternal}) may be
written also in the form
\begin{gather}
\label{Ese}
S^\mn=\ep^{\mn\rho\lambda}p_\rho\theta^{ai}\sigma_{\lambda\,
a\db}\bar \theta\ud{\db}{i}\ \ .
\end{gather}

\section{The quantum algebra}

In order to construct a  quantum theory on the superspace, related
to the classical system of the previous section or to any other
supersymmetric system of interest formulated on this superspace
one should be able to identify a set of quantum operators
satisfying the  corresponding algebra. For the superparticle this
algebra is identified by applying the quantization rules to the
Dirac algebra discussed above. For other systems it should be a
suitable covariant generalization. In the functional space of all
wave functions defined on the $N$ superspace the standard position
operators $X_\mu$, $\Theta^a$, and $\bar\Theta^\da$ act
multiplicatively and together with
\begin{gather}
P_\mu=-i\p_\mu\\
\Pi_{ai}=-i\p_{ai}\qquad \bar\Pi^i_\da=-i\bar\p^i_\da\\
Q_{ai}=\Pi_{ai}-iP_\mu\sigma\ud{\mu}{a\da}\bar\Theta_i^\da\\
\bar Q^i_\da=-\bar\Pi^i_\da+iP_\mu\sigma\ud{\mu}{a\da}\Theta^{ai}\
\ .
\end{gather}
give   a representation of the super Poincar{\'e} algebra without
central charges. \cite{SalStr,WesBag}. The covariant derivatives
are defined as
\begin{gather}
D_a=\p_a+i\sigma\ud{\mu}{a\db}\bar\Theta^\db\p_\mu=i\Pi_a-P_\mu\sigma\ud{\mu}{a\db}\bar\Theta^\db\label{CovDer1}\\
\bar
D_\db=-\bar\p_\db-i\Theta^a\sigma\ud{\mu}{a\db}\p_\mu=-i\bar\Pi_\db+P_\mu\Theta^a\sigma\ud{\mu}{a\db}\label{CovDer2}
\end{gather}
The total angular momentum is given by,
\begin{gather}
J_\mn=L_\mn+S_\mn\\
L_\mn=X_\mu P_\nu-X_\mu P_\nu\ \ , \ \ S_\mn=-\frac{1}{4}
(\Theta^i\sigma_\mn\Pi_i+\bar\Pi^i\bar\sigma_\mn\bar\Theta_i)
\label{IntAM}
\end{gather}
This representation is reducible and  does not correspond to the
quantization of the superparticle. Let us introduce another set of
quantum operators $\hat X^\mu$, $\hat \Theta^{ai}$ and
$\bar\Theta\ud{\da}{i}$  and their corresponding momentum
operators acting on the superfields in such a way that the
corresponding representation of the super Poincar{\'e} algebra is
irreducible. We propose that these operators should satisfy the
quantum algebra given by
\begin{gather}
\con{\hat P_\mu,\hat\Theta^{ai}}=\con{\hat P_\mu,\hat
{\bar\Theta}\ud{\da}{i}}= \cor{\hat\Theta^{ai},\hat\Theta^{bj}}
=\cor{\hat{\bar\Theta}\ud{\da}{i},
\hat{\bar\Theta}\ud{\db}{j}}=0\label{conPrime}\\
\con{\hat X^\mu,\hat P^\nu}=i\eta^\mn\\
\cor{\hat\Theta^{ai},\hat{\bar\Theta}\ud{\da}{j}}=\frac{-1}{2\hat
P^2}
\hat P_\mu\bar \sigma^{\mu\,\da a}\delta\ud{i}{j}\label{con1Prime}\\
\con{ \hat X^\mu, \hat\Theta^{ai}}=\frac{i}{2\hat P^2}
\hat P_\nu \sigma^{\nu\,\da a} \hat \Theta^{bi} \sigma\ud{\mu}{b\da} \label{XTheta}\\
\con{ \hat X^\mu,\hat{\bar\Theta}\ud{\da}{i}}=\frac{i}{2\hat
P^2}\hat P_\nu \bar \sigma^{\nu\,\da a}
\hat{\bar\Theta}\ud{\db}{i}\sigma\ud{\mu}{a\db}\label{XBarTheta}\\
\con{ \hat X^\mu, \hat X^\nu}=\frac{-i\hat S^\mn}{\hat P^2}\ \ ,
\label{ConCoord}
\end{gather}
where $\hat S^\mn$ is the internal angular momentum operator in in
the subspace where we are representing the algebra and reduces to
the form (\ref{IntAM})(in terms of the new operators) for the
representation corresponding to superspin 0 supermultiplets. This
algebra indeed generalizes  the quantum algebra of the massive
superparticle obtained by applying the quantization rules to the
classical Dirac described in the previous section. As stated
before to perform the quantization of the system one should
exhibit a set of operators and a space of wave functions for which
the quantum algebra above holds. The related problem was solved
for $N=1$ in \cite{ResHS2006} using the properties of the
operators \cite{GatGriRoeSie,SieG1981,RitSok} which project
superfields to the chiral, anti-chiral and tensorial
supermultiplets which carry the irreducible representations of the
supersymmetry algebra present in the superfields in this case.
Following this approach let us consider the operators $\Proy_\GG$
acting on the superfields which project to the irreducible
representations of the supersymmetry in the extended case. As we
show below the projected operators
\begin{gather}
\hat X^\mu=X^\mu_\GG\equiv \Proy_\GG X^\mu \Proy_\GG\label{Op1} \\
\hat\Theta^{ai}=\Theta^{ai}_\GG\equiv \Proy_\GG\Theta^a \Proy_\GG\label{Op2}\\
\hat{\bar\Theta}\ud{\da}{i}={\bar\Theta}_{G\,i}^\da\equiv
\Proy_\GG\bar\Theta\ud{\da}{i}
\Proy_\GG\label{Op3}\\
\hat P_\mu=P^\mu_\GG\equiv \Proy_\GG P^\mu\Proy_\GG \label{Op4}\ \
,
\end{gather}
satisfy the desired algebra with the internal angular momentum
given by,
\begin{gather}
\hat S^\mn=S^\mn_\GG\equiv \Proy_\GG S^\mn
\Proy_\GG\label{TotalS1}
\end{gather}
For the particular case in which $\Proy_\GG$ projects to the
superspin zero representation we obtain the quantization of the
massive superparticle. The explicit expressions for $\Proy_\GG$ in
$D=4$ for arbitrary $N$ were obtained in Ref. \cite{SieG1981}. In
terms of the covariant derivatives they are of the general form
\begin{gather}
\Proy_\GG \propto D^{[q}{\bar D}^{2N}D^{p]}
\end{gather}
where the square brackets mean that the corresponding indices are
contracted and $q+p=2N$. They satisfy,
\begin{gather}
\Pg D_{ai}\Pg=\Pg \bar D\du{\da}{i}\Pg=0\label{IrrCondition}\\
\con{\Pg,Q_{ai}}=\con{\Pg,\bar
Q\du{\da}{i}}=\con{J^\mn,\Pg}=\con{P_\mu,\Pg}=0
\label{CommCondition}
\end{gather}

A direct computation of the quantum algebra using the expressions
for the projectors as was done for the $N=1$ case is not practical
in this case. Instead we present our general result in the
following theorem valid with convenient modifications also for $D> 4$\\
\textbf{Theorem}\\
\textit{Let $\Proy_\GG$ be a projector operator that satisfy
(\ref{IrrCondition}) and (\ref{CommCondition}). Then the set of
operators (\ref{Op1}-\ref{Op3}) satisfy relations
(\ref{conPrime}-\ref{ConCoord}) and $S^\mn_\GG$ may be expressed
in the form
\begin{gather}
S^\mn_\GG=\tilde S_\GG^\mn+ W^\mn_\GG\label{InterAngular}
\end{gather}}\\
with
\begin{gather}
\label{eq:STDD}
\tilde S^\mn_\GG=-\frac{1}{4}\left(\Theta_\GG\sigma^\mn\Pi_\GG+\bar\Pi_\GG\bar\sigma^\mn\bar\Theta_\GG\right)\\
\label{eq:WDD}
W^\mn_\GG=\frac{P_\alpha}{4P^2}\ep^{\mn\alpha\lambda}\Pg\bar
D^i\bar\sigma_\lambda D_i\Pg
\end{gather}\\
\textbf{Proof}\\
To obtain this result we consider all the  operators corresponding
to the generators of the super Poincar{\'e} algebra projected as
(\ref{Op1}-\ref{Op4}). Since the projectors commute with the
generators the projected generators also satisfy the super
Poincar{\'e} algebra. In particular let us take the supercharges
$Q_{ai}^\GG$ and $\bar Q\du{\GG\,\da}{i}$.  The first step is to
note that equations (\ref{IrrCondition}) are equivalent to
\begin{gather}
\Pi_{ai}^\GG=-i P_\mu\sigma\ud{\mu}{a\da}{\bar\Theta}^\da_{\GG i}\\
{\bar\Pi}_\da^{\GG i}=-i P_\mu\sigma\ud{\mu}{a\da}\Theta^{ai}_\GG\
\ .
\end{gather}
Here and in what follow we use that $\con{P_\mu,\Pg}=0$ and we
write $P_\mu$ instead of $P^\mu_\GG$.  We have then,
\begin{gather}
Q_{ai}^\GG=-2iP_\mu \sigma\ud{\mu}{a\db}\bar\Theta\dud{\GG}{\db}{i}\\
\bar Q\du{\GG\,\da}{i}=2iP_\mu\sigma\ud{\mu}{a\da}\Theta^{ai}_\GG
\ \ ,
\end{gather}
and so we may write $\Theta^{ai}$ and $\bar\Theta\ud{\db}{i}$ in
terms of the charges
\begin{gather}
\label{eq:ThetaQbar}
\Theta^{ai}_\GG=\frac{i}{4P^2}P_\mu\bar\sigma^{\mu\,\db a}\bar Q\udu{\GG}{\db}{i}\\
\bar
\Theta\dud{\GG}{\db}{i}=\frac{-i}{4P^2}P_\mu\bar\sigma^{\mu\,\db
b}Q^G_{bi} \ \ .
\end{gather}
Using the supersymmetry algebra for $Q_{ai}^\GG$ and $\bar
Q\udu{\GG}{\db}{i}$, one readily shows that relations
(\ref{conPrime}) to (\ref{con1Prime}) hold. To see that relations
(\ref{XTheta}) and (\ref{XBarTheta}) also hold we need only to
note that
\begin{gather}
\con{X^\mu_\GG,Q_{ai}^\GG}=-\sigma\ud{\mu}{a\da}\bar\Theta\dud{\GG}{\da}{i}\\
\con{X^\mu_\GG,\bar
Q\udu{\GG}{\da}{i}}=\sigma\ud{\mu}{a\da}\Theta\du{\GG}{ai}
\end{gather}
Finally we show the non commutativity of $X^\nu_\GG$. We first
proof equation (\ref{InterAngular}). Writing $\Pi_{ai}$ and
$\Theta^{ai}$ in terms of the covariant derivatives and the super
charges,
\begin{gather}
\Pi_{ai}=\frac{1}{2}(Q_{ai}-iD_{ai})\\
\Theta^{ai}=\frac{-1}{2P^2}P_\nu\bar\sigma^{\nu\,\db a}(\bar
D\du{\db}{i}-i\bar Q\du{\db}{i})
\end{gather}
and using that $\Pg D_{ai}\Pg=0$ implies also $\Pg Q_{bj}
D_{ai}\Pg=0$ we see that
\begin{gather}
S^\mn_\GG=\Pg S^\mn \Pg=\frac{-1}{4}\Pg\left(\Theta^i\sigma_\mn\Pi_i+\bar\Pi^i\bar\sigma_\mn\bar\Theta_i)\right)\Pg=\nonumber \\
\frac{P_\alpha}{8P^2}\Pg\left((\bar D^{i}-i\bar Q^{i})\bar\sigma^{\alpha} \sigma_\mn(Q_{i}-iD_{i})+\textrm{c.c.}\right)\Pg=\nonumber\\
\frac{-iP_\alpha}{8P^2}\Pg\left(\bar D^i
\bar\sigma^{\alpha}\sigma_\mn D_i\right)\Pg+
\frac{-iP_\alpha}{8P^2}\Pg\left(\bar Q^i
\bar\sigma^{\alpha}\sigma_\mn Q_i\right)\Pg+\textrm{c.c.}
\end{gather}
The first term is (\ref{eq:WDD}). For the second term we  use
equation (\ref{eq:ThetaQbar}) and the relation $Q^\GG=2\Pi^\GG$ to
obtain the right side of (\ref{eq:STDD}). This prove equation
(\ref{InterAngular}) for the projected angular momentum operator.
We note that from (\ref{InterAngular}), Pryce condition \cite{Pry}
$P_\mu S^\mn=0$ may be readily deduced. Using this and the algebra
of $J^\mn$ one obtains
\begin{gather}
\con{J^\mn,X^\lambda_\GG}=-iX^\mu_\GG \eta^{\nu\lambda}+iX^\nu_\GG\eta^{\mu\lambda}\label{eq:readily}\\
\con{S^\mn_\GG,X^\lambda_\GG}=-iX^\mu_\GG
\eta^{\nu\lambda}+iX^\nu_\GG\eta^{\mu\lambda}-
\con{X^\mu_\GG P^\nu-X^\nu_\GG P^\mu,X^\lambda_\GG}\\
\con{S^\mn_\GG,X^\lambda_\GG}=-\con{X^\mu_\GG,X^\lambda_\GG}P^\nu+\con{X^\nu_\GG,X^\lambda_\GG}P^\mu\\
0=\con{P_\mu S^\mn_\GG,X^\lambda_\GG}=-iS^{\lambda\nu}+P_\mu\con{S^\mn_\GG,X^\lambda_\GG}\\
iS^{\lambda\nu}=-P_\mu
\con{X^\mu_\GG,X^\lambda_\GG}P^\nu+P_\mu\con{X^\nu_\GG,X^\lambda_\GG}P^\mu
\end{gather}
Note that the Jacobi identity  implies that $P_\mu$ commutes with
$\con{X^\alpha_\GG,X^\beta_\GG}$.  In the last equation
multiplying by $P_\nu$ and doing the sum  it is shown that
$P_\nu\con{X^\nu_\GG,X^\lambda}=0$. From here  the result
\begin{gather}
\con{X^\mu_\GG,X^\nu_\GG}=-iS_\GG^\mn/P^2
\end{gather}
is obtained.
\section{The Pauli-Lubanski vector and $W^\mn$}
Let us see now that the $W^\mn$ term  of the internal angular
momentum operator (see (\ref{InterAngular})) is closely related to
the super Pauli-Lubanski  vector and fix the superspin of the
representation. Defining the Pauli-Lubanski vector in the
representation space projected by $\Proy_\GG$ in the usual form
\begin{gather}
W_\mu^G=\Proy_\GG W_\mu\Proy_\GG=\frac{1}{2}\ep_{\mn\alpha\beta}
P^\nu_\GG J^{\alpha\beta}_\GG+ \frac{1}{8}{\bar
Q}^i_\GG\bar\sigma_\mu Q_i^G\ \ ,
\end{gather}
where $J^{\mn}_\GG=L^\mn_\GG+S^\mn_\GG$ and
\begin{gather}
S^\mn_\GG=\tilde S^\mn_\GG+ W^\mn_\GG\\
W^\mn_\GG=\frac{-P_\alpha}{4P^2}\ep^{\mn\alpha\lambda}\Proy_\GG
D_i\sigma_\lambda\bar D^i\Proy_\GG
\end{gather}
one can show that,
\begin{gather}
\frac{1}{2}\ep_{\mn\alpha\beta} P^\nu_\GG (L^\mn_\GG+\tilde
S^\mn_\GG)+ \frac{1}{8}{\bar Q}^i_\GG\bar\sigma_\mu Q_i^G=0\ \ .
\end{gather}
Then, the only non vanishing contribution to the super
Pauli-Lubanski four vector comes  from the $W^\mn$ term and one
can write
\begin{gather}
W_\mu^G=\frac{1}{2}\ep_{\mn\alpha\beta} P^\nu_\GG
W^{\alpha\beta}_\GG
\end{gather}
or equivalently
\begin{gather}
W^{\mu\nu}_\GG=\ep^{\mn\alpha\beta} P_{\GG\alpha} W_\beta^G\ \ .
\end{gather}
This shows directly that in the quantum theory of the
superparticle for which one should have $W^{\mu\nu}_\GG=0$,
$W_\mu^G$ also vanishes an so the corresponding super multiplet
has superspin 0. Supermultiplets with higher values of the
superspin appear in the other cases.

\section{Central charges and superprojectors}
\label{sec:CentralCharges} Let us now turn to the construction of
the quantum algebra when there are central charges in the
supersymmetry algebra.  In what follows we consider the $N=2$
superspace. Other cases may be discussed along the same lines.

The supersymmetry algebra is given  by
\begin{gather}
\cor{Q_{ai},Q_{bj}}=2i\ep_{ab}\ep_{ij}Z\qquad \cor{Q_{ai},
\bar{Q}\du{\da}{j}}=2\delta\du{i}{j}\sigma\ud{\mu}{a\da}P_\mu
\end{gather}
Superspace  coordinates are now $(x^\mu,z,\bar
z,\theta^{ai},\bar\theta\ud{\da}{i})$ an the algebra is
represented on the superfields by the operators
\begin{gather}
Z=-i\frac{\p}{\p z}\qquad \bar Z=-i\frac{\p}{\p \bar z}\\
\Pi_{ai}=-i\p_{ai}\qquad \bar \Pi\du{\da}{i}=-i\bar \p\du{\da}{i}
Q_ai=\Pi_{ai}-iP_\mu\sigma\ud{\mu}{a\da}\bar\Theta\ud{\da}{i}-\ep_{ab}\ep_{ij} Z\Theta^{bj}\label{eq:Qcentral}\\
\bar Q\du{\da}{i}=-\bar\Pi\du{\da}{i}+i\Theta^{ai}P_\mu\sigma\ud{\mu}{a\da}-\ep_{\da\db}\ep^{ij}\bar Z\bar \Theta\ud{\db}{j}\\
Q_{ai}=-i\p_{ai}-\p_\mu\sigma\ud{\mu}{a\da}\bar\theta_i^\da+i\ep_{ab}\ep_{ij}\theta^{bj}\p_z\label{eq:QBARcentral}\\
\end{gather}

As a guide for the construction of the quantum algebra let us
consider the dynamics of the superparticle in this superspace. The
action for this system  was proposed some time ago by Azcarraga
and Lukierski \cite{AzLuk82}. There are three simple invariants
that can be build in terms of the coordinates, $w^\mu w_\mu$ as
before, $(\ep_{ab}\ep_{ij}\theta^{ai}\dot\theta^{bj}+\dot z)$ and
its complex conjugate. Therefore a general action invariant under
the supersymmetry algebra is
\begin{gather}
S=\int\left( \frac{1}{2}(e^{-1}w^\mu
w_\mu-em^2)+\ell(\theta^{ai}\dot\theta^{bj}+\dot z)+\ell^*(\dot
{\bar
z}-\bar\theta\ud{\da}{i}\dot{\bar\theta}\du{\da}{i})\right)d\tau
\end{gather}
Here $\ell$ is an arbitrary parameter with mass units. Generalized
momenta are given by
\begin{gather}
p_\mu=e^{-1}\omega_\mu\qquad \pi_z=\ell\qquad \pi_{\bar z}=\ell^*\\
\pi_{ai}=-ip_\mu\sigma\ud{\mu}{a\db}\bar\theta\ud{\db}{i}-\pi_z\theta_{ai}\
\ \ \bar
\pi\du{\da}{i}=-ip_\mu\sigma\ud{\mu}{a\da}\theta^{ai}-\bar\theta\du{\da}{i}\pi_{\bar
z}
\end{gather}
The constraints are now
\begin{gather}
\pi_e=0\rightarrow p^2+m^2=0\\
d_{ai}=\pi_{ai}+ip_\mu\sigma\ud{\mu}{a\db}\bar\theta\ud{\db}{i}+\pi_z\theta_{ai}\\
\bar d\du{\da}{i}=\bar
\pi\du{\da}{i}+ip_\mu\sigma\ud{\mu}{a\da}\theta^{ai}-\bar\theta\du{\da}{i}\pi_{\bar
z}
\end{gather}
And satisfy:
\begin{gather}
\label{Vin1} \cor{d_{ai},d_{bj}}=-2\ep_{ij}\ep_{ab}\pi_z \qquad
\cor{\bar d\du{\da}{i},\bar d\du{\db}{j}}=2\ep^{ij}\ep_{\da\db}\pi_{\bar z}\\
\cor{d_{ai},\bar
d\du{\da}{j}}=-2i\delta\du{i}{j}p_\mu\sigma\ud{\mu}{a\da}
\label{Vin2}
\end{gather}
The nature of the constraints depends on the value of $\ell$.We
consider first the case $|\ell|\neq m$. Then, all the constraints
are second class. Introducing ${\hat C}$ the inverse of the
constraints matrix defined by (\ref{Vin1}-\ref{Vin2}) we have,
\begin{gather}
{\hat C}^{abij}=\frac{1}{2\Delta}\ep^{ab}\ep^{ij}\pi_{\bar z}\\
{\hat C}\ud{a\da i}{j}=\frac{1}{2i\Delta}p_\mu\bar\sigma^{\mu\,\da a}\delta\ud{i}{j}\\
\Delta=p^2+\pi_z\pi_{\bar z}
\end{gather}

The Dirac brackets are then obtained as
\begin{gather}
\dir{\theta^{ai},\theta^{bj}}=-{\hat C}^{abij}\qquad\cor{\bar\theta\ud{\da}{i},\bar\theta\ud{\db}{j}}=-\bar {\hat C}\ud{\da\db}{ij}\\
\dir{\theta^{ai},x^\mu}=-{\hat
C}^{acik}i\sigma\ud{\mu}{c\da}\bar\theta\ud{\da}{k}-{\hat
C}\ud{a\da i}{k}i\sigma\ud{\mu}{b\da}\theta^{bk}\\
\dir{x^\mu,x^\nu}=-\frac{S^\mn}{p^2}
\end{gather}
where $S^\mn$ is given by (\ref{ClassicalInternal}) which in this
case is not equal to (\ref{Ese}).

The quantization of this system may be done on the space of
superfields using the projectors approach. The straightforward
quantization of the classical algebra leads to the following
quantum algebra
\begin{gather}
\label{Qacc}
\cor{\hat\Theta^{ai},\hat\Theta^{bj}}=\frac{i}{2\Delta}\ep^{ab}\ep^{ij}\hat \Pi_{\bar z}\\
\cor{\hat{\bar\Theta}\ud{\da}{i},\hat{\bar \Theta}\ud{\db}{j}}=\frac{-i}{2\Delta}\ep^{\da\db}\ep_{ij}\hat\Pi_z\\
\con{\hat\Theta^{ai},\hat X^\mu}=\frac{1}{2\Delta}\ep^{ab}\ep^{ij}\ell^*\sigma\ud{\mu}{b\db}\hat{\bar\Theta}\ud{\db}{j}+\frac{i}{2\Delta}P_\mu P_\nu\bar\sigma^{\nu\,a\da}\sigma\ud{\mu}{b\da}\hat\Theta^b\\
\con{\hat X^\mu, \hat X^\nu}=\frac{-i \hat S^\mn}{P^2}
\end{gather}
To find a representation of this quantum algebra we use the
superprojectors method discussed above.The superprojector operator
$\Proy_\GG$ should commute with all the generators of the super
Poincar{\'e} algebra and  satisfies
\begin{gather}
\Proy_\GG D_{ai}\Proy_\GG=\Proy_\GG\bar D\du{\da}{i}\Proy_\GG=0\\
D_ai=i\Pi_{ai}-P_\mu\sigma\ud{\mu}{a\da}\bar\Theta\ud{\da}{i}+i\ep_{ab}\ep_{ij} Z\Theta^{bj}\label{eq:Dcentral}\\
\bar
D\du{\da}{i}=-i\bar\Pi\du{\da}{i}+\Theta^{ai}P_\mu\sigma\ud{\mu}{a\da}+i\ep_{\da\db}\ep^{ij}\bar
Z\bar \Theta\ud{\db}{j} \label{eq:DBARcentral}
\end{gather}
Using the same techniques as for the case without central charges
we can show that using the operators
\begin{gather}
\hat{X}^\mu=\Proy_\GG X^\mu \Proy_\GG\\
\hat{\Theta}^\mu=\Proy_\GG \Theta^\mu \Proy_\GG\\
\hat{\bar\Theta}^\mu=\Proy_\GG \bar\Theta^\mu \Proy_\GG
\end{gather}
realize the quantum algebra (\ref{Qacc}). The first step is to
note that using (\ref{eq:Qcentral}, \ref{eq:QBARcentral},
\ref{eq:Dcentral}, \ref{eq:DBARcentral}), $\theta^{ai}$ and
$\bar{\theta}\ud{\da}{i}$ can be written in terms of the
supercharges and the covariant derivatives. In a matrix form:
\begin{gather}
\mat{\Theta^{ai}\\ \bar\Theta\ud{\da}{i}}=\mat{\frac{-1}{2\Delta}\ep^{ab}\ep^{ij}\bar Z & \frac{1}{2i\Delta}P_\mu\bar\sigma^{\mu\,\db a}\delta\ud{i}{j}\\
\frac{1}{2i\Delta}P_\mu\bar \sigma^{\mu\,\da a}\delta\du{i}{j} &
\frac{1}{2\Delta}\ep^{\da\db}\ep_{ij}Z} \mat{Q_{bj}+iD_{bj}\\
-\bar Q\du{\db}{j}-i\bar D\du{\db}{j}}
\end{gather}
This implies, in particular, that:
\begin{gather}
\hat\Theta^{ai}=\frac{-1}{2\Delta}\ep^{ab}\ep^{ij}\bar Z\hat Q_{bj}-\frac{1}{2i\Delta}P_\mu\bar\sigma^{\mu\,\da a}\hat{\bar Q}\du{\da}{i}\\
\hat{\bar \Theta}\ud{\da}{i}=\frac{1}{2\Delta}P_\mu\bar
\sigma^{\mu\, \da a}\hat
Q_{ai}-\frac{1}{2\Delta}\ep^{\da\db}\ep_{ij} Z\hat Q\du{\db}{j}
\end{gather}
With these formulas at hand is easy to check  all commutators in
(\ref{Qacc} except Pryce relation. To prove this  note that  $\hat
S^\mn$ may be rewritten  exactly in the same form as in the case
without central charges
\begin{gather}
\hat S^\mn=\frac{-1}{4}\left(\hat{\Theta}^i\sigma_\mn\hat
\Pi_i+\hat{\bar\Pi}^i\bar
\sigma_\mn\bar\Theta_i\right)-\frac{-iP_\alpha}{8P^2}\Proy_{\GG}\bar
D^i\bar \sigma_\mn D_i\Proy_\GG
\end{gather}
Now the arguments that follow equation \ref{eq:readily} remain
unchanged and we conclude again
\begin{gather}
\con{\hat X^\mu,\hat X^\mu}=\frac{-i\hat S^\mn}{P^2}
\end{gather}
\section{The kappa symmetric case}
The set of constraints imply
\begin{gather}
\ell\bar\pi\du{\da}{i}-ip_\mu\sigma\ud{\mu}{a\da}\pi^{ai}-(p^2+|\ell|^2)\bar\theta\du{\da}{i}=0
\end{gather}
If $|\ell|=m$ this constraints are first class and indeed the
system is $\kappa$-symmetric. Define now,
\begin{gather}
C\du{a}{i}=\ell^*\pi_{ai}+ip_\mu\sigma\ud{\mu}{a\da}\bar \pi\ud{\da}{i}\\
\bar C\du{\da}{i}=\ell\bar
pi\du{\da}{i}-p_\mu\sigma\ud{\mu}{a\da}\pi^{ai}
\end{gather}
Following \cite{AzLuk82} we have a set of independent first class
constraints,
\begin{gather}
\phi=p^2+m^2=0\qquad \pi_z-\ell=0\qquad \pi_{\bar z}-\ell^*=0\\
C\du{a}{2}=0\qquad \bar C\du{\da}{2}=0
\end{gather}
together with a set of second class constraints
\begin{gather}
d_{a1}\qquad \bar d\du{\da}{1}\ .
\end{gather}

To make the transition to the quantum theory we can split the
second class constraints into a set of first class constraints and
some gauge fixing conditions or we may rely on Dirac's procedure
and use super projectors. Of course, both mechanisms should  give
the same answers. The first displays more clearly the particle
content of the theory while the second allows
to represent directly position operators as self adjoint operators in a Hilbert space.\\

To follow the first approach we should specify an admissible set
of first class constraints. For example we can take $\bar
d\du{\da}{2},\quad \bar C\du{\da}{2},\quad C\du{a}{2}$. The
quantum mechanics of this system should  then be formulated in
terms of a superfield
$V((x^\mu,\theta^{ai},\bar\theta\ud{\da}{i},z)$ that satisfy the
equations:
\begin{gather}
(P^2+m^2)V=0\qquad \bar D\du{\da}{1}V=0\qquad -i\p_zV=\ell V\qquad i\p_{\bar z}V=\ell^*V\\
\ell \bar \p_\da^2V=\sigma\ud{\mu}{a\da}\p_\mu\p^{a2}V\\
\ell^*\p_{ai}V=-\sigma\ud{\mu}{a\da}\p_\mu\bar \p\ud{\da}{i}V\ .
\end{gather}

Let us consider now the method of superprojectors. Dirac's
procedure for this system has  been carried out with some detail
in \cite{AzLuk82} so we need only to quote the results. First note
that the Dirac bracket of $\theta^{a2}$ and
$\bar\theta\ud{\da}{i}$ with anything remains unchanged because
they commute with the constraints ($d_{a2}$ and $\bar
d\du{\da}{2}$). The algebra of the $\pi$'s can always be deduced
from the rest so the independent degrees of freedom are encoded in
the variables $(x^\mu,p_\nu,\theta^{a1},\bar\theta\ud{\da}{1})$
and the algebra of this variables is exactly the same as the $N=1$
case and is given by equations (\ref{con1}-\ref{con2}) with
$i,j=1$ also the $S^\mn$ piece
is given by (\ref{Ese}) but only keeping the $i=1$ term.\\
Quantization now follows exactly the same rules that the $N=1$
case but the projectors are constructed from the following
covariant derivatives
\begin{gather}
D_{ai}=\p_{ai}+i\sigma\ud{\mu}{a\db}\bar\Theta\ud{\db}{i}\p_\mu+m\ep_{ab}\ep_{ij}\theta^{aj} \label{eq:CovCentralDer1}\\
\bar
D\du{\db}{i}=-\bar\p\du{\db}{i}-i\Theta^{ai}\sigma\ud{\mu}{a\db}\p_\mu+m\ep_{\da\db}\ep^{ij}\theta\ud{\da}{j}
\label{eq:CovCentralDer2}
\end{gather}

\section{Quantization of massless superparticles\label{sec:masless}}
The discussion of the previous section suggests a way to treat
another interesting problem, that of the quantization of massless
superparticles in $D=4,6,8$. It is known that for massless
particles it is  difficult to define position operators. Here this
is reflected in the fact that for $m=0$ the constraints
(\ref{eq:d1}-\ref{eq:d2}) are a mixture of first and second class
constraints. Also it is not easy to disentangle the first and
second class constraints in a Lorentz covariant an irreducible way
or to find a Lorentz covariant gauge fixing condition. Instead we
note that the set of first and second class constraints
\begin{gather}
p^2,d_a,\bar d_\da\label{eq:secondclass}
\end{gather}
is equivalent to a system with only first class constraints
\begin{gather}
p^2, \bar d_a, \phi^a\equiv S^{\mu\,ab}P_\mu d_b
\label{eq:firstclass}
\end{gather}
That is the set (\ref{eq:secondclass}) is obtained from the set
(\ref{eq:firstclass}) upon gauge fixing. To show this we need to
see that if $d_a$ is any field satisfying $p_\mu S^{\mu\,ab}d_b=0$
then there is a gauge transformation generated by $\bar d_\da$
that sets $d_a=0$. Now the solution of $p_\mu S^{\mu\,ab}d_b=0$ is
$d_b=S\ud{\mu}{ba}\xi^\da$, so the only thing we have to do is
make a gauge transformation generated by $\xi^\da \bar d_\da$. Now
quantization is straightforward. The superfield satisfies
\begin{gather}
\bar D_\da V=0\\
S^{\mu\,\da b}D_b\p_\mu V=0
\end{gather}
This procedure also work nicely for the $D=10$, $IIB$
superparticle. In that case we have two Majorana-Weyl
$\theta^{ai}$ of the same chirality that can be recast in a single
$\theta^a$, $\bar \theta^\da$ Weyl, but not Majorana fermion. In
this language $D=10$, IIB superspace can be treated as before, so
that the above ecuations can also be considered as a quantization
of the IIB superparticle and are a complex extension of the
linearized IIB supergravity.
\section{Conclusion}
In this paper we generalized the construction of the quantum
algebra of the $N=1$, $D=4$ superspace without central charges for
an arbitrary number $N$ of supersymmetric charges. This was done
using the properties of the superprojectors which project to the
irreducible supersymmetric multiplets. Here again the most
remarkable aspect of the algebra is related to the non trivial
commutator of the position operators which is proportional to the
internal angular momentum operator. As another interesting point
we show that the quantum internal angular momentum splits in the
general case in two parts, one with the form suggested by the
quantization of the superparticle and a second term which also
allows to represent express the super Pauli-Lubanski vector. Our
construction gives in particular a complete covariant solution of
the quantization of the massive superparticle in terms of the
superspin 0 supermultiplet, but  includes also as for $N=1$, the
quantum algebra of other different systems with higher values of
the superspin. The identification of the corresponding
pseudoclassical actions may bring some clues on the physics of ten
dimensional superspace.

We also applied the same methods to discuss the quantum algebras
associated to the $N=2$, $D=4$ superspace with central charges and
their relation with the corresponding superparticle.  For the
$/kappa$-symmetric case when the central charge and the mass are
are equal we discuss the reduction to the physical degrees of
freedom of the  superparticle by a reinterpretation of some of the
second class constraints as gauge fixing conditions. We also
mention how this approach may also be applied to massless
particles and the $D=10$, $IIB$ superparticle.

Our methods could be generalized to other dimensions in particular
$D=10$ \cite{ResHS2006b} where they may provide a new approach for
the covariant quantization of supersymmetric systems.

\section{Acknowledgments} This work was supported
by Did-Usb grants Gid-30 and Gid-11 and by Fonacit grant
G-2001000712.


\begin{thebibliography}{long}
\bibitem{BatF1983} I.A.Batalin and E.S.Fradkin, Phys Lett B \textbf{122},  157,(1983).
\bibitem{RS} A. Restuccia and J. Stephany, Phys Lett B \textbf{305}, 348, (1993); Phys Lett B \textbf {43}, 147, (1995).
\bibitem{BizS1996} C. Bizdadea and S.O. Saliu, Nucl Phys B \textbf{469}, 302, (1996).
\bibitem{SeiW199}N. Seiberg and E. Witten, JHEP \textbf{9909},032,(1999).
\bibitem{ResHS2006}N. Hatcher, A. Restuccia and  J. Stephany, Phys Rev D, \textbf{73} 046008, (2006) /hep-th 0511066.
\bibitem{Hat} Nicol{\'a}s Hatcher, \textit{Phd} thesis, Universidad Sim{\'o}n Bol{\'\i}var 2006, unpublished.
\bibitem{Cas2} R. Casalbuoni, Nuov Cim A \textbf{33}, 389, (1976).
\bibitem{Pry} M. H. L. Pryce, Proc Royal Soc of Lon A \textbf{195}, 62, (1948).
\bibitem{SalStr} A. Salam and J. Strathdee, Phys Rev D \textbf{11}, 1521, (1975).
\bibitem{WesBag} J. Wess and Bagger \textit{``Supersymmetry and supergravity"}, Princeton Series in Physics, (1982).
\bibitem{GatGriRoeSie} S.J. Gates, M.T. Grisaru, M. Rocek and W. Siegel \textit{"One thousand and one lesons in supersymmetry"}, Frontiers in Physics Vol 58, (Benjamin, New York, 1983).
\bibitem{SieG1981} W. Siegel, S.J. Gates,Jr.  Nucl Phys B \textbf{189}, 295, (1981).
\bibitem{RitSok} V. Rittenberg and E. Sokatchev, Nuc Phys B, \textbf{193},  477, (1981).
\bibitem{ResHS2006b}  Nicol{\'a}s Hatcher, A. Restuccia and  J. Stephany, hep-th/0604035.
\bibitem{AzLuk82} J.A.de Azc{\'a}rraga and J.Lukierski, Phys Lett B, textbf{113},  170,(1982)
\end{thebibliography}
\end{document}